\documentstyle[12pt,aasms4,epsfig,rotate]{article}
\newcommand{\etal}{ {\it et al.}}

\newcommand{\msol}{M$_{\sun}$}

\begin{document}

\title{Spectral State Transitions in Aql X-1: Evidence for `Propeller' Effects}
\author{S. N. Zhang$^{1,2}$, W.~Yu$^{1,4}$, W. Zhang$^{3}$}
\affil{$^{1}$ES-84, NASA/Marshall Space Flight Center, Huntsville, AL 35812, USA\\
$^{2}$Universities Space Research Association\\
$^{3}$NASA/GSFC, Code 662, Greenbelt, MD 20771\\
$^{4}$LCRHEA, Institute of High Energy Physics, Beijing, China\\
}

\vspace{3cm}
\begin{center}
\end{center}

\abstract{
Aql X-1 is a soft X-ray transient source and emits type I X-ray
bursts.  A spectral state transition was observed with RXTE during its
outburst decay in February and March 1997. Its 10-30 keV and 5-10 keV
count rate ratio increased suddenly when its luminosity was between
(4-12) E35 ergs/s, assuming a 2.5 kpc distance. Spectral fitting with a
model composed of a blackbody and a power-law components showed that
its blackbody component decreased and the power-law became much harder
significantly and simultaneously. We interpret this transition as
caused by the centrifugal barrier, or commonly known as the `propeller'
effect. We thus infer that the magnetic field strength of the neutron
star is around 1E8 Gauss, if the neutron star spin period is 1.8 ms.
Similarly we infer the neutron star magnetic field strength in another
soft X-ray transient Cen X-4 is about 2E9 Gauss. We also
propose a unified scheme for spectral state transitions in soft X-ray
transients, from soft high state to hard low state and further to
quiescent state.  With this scheme accretion onto neutron star may take place
even during the propeller regime.  }


\section{Introduction}

Aql X-1 is a soft X-ray transient (SXT) with rather frequent outbursts
(see for example, Priedhorsky \& Terrell 1984; Kitamoto \etal\ 1993;
Harmon \etal\ 1996).  It is believed to contain a weakly magnetized
neutron star (NS) as indicated by its type I X-ray bursts
(Koyama \etal\ 1980; Czerny, Czerny \& Grindlay 1987). From the
observed 549 Hz oscillations of X-ray flux during one of its X-ray
bursts, its NS is believed to be spinning at a period ($P$) of either
1.8 or 3.6 ms (White \& Zhang 1997, Zhang \etal\ 1997), thus among the
fastest spinning NS known. The strength of its magnetic field ($B$) has
never been determined from observations, although believed to be quite
weak, as inferred from the lack of any detectable pulsed X-ray flux in
its persistent emission.

Aql X-1 has been observed to exhibit three types of X-ray spectral
states. When its luminosity is above $\sim$10$^{36}$ ergs/s (assuming
a distance of 2.5 kpc), its spectrum is very soft (Czerny, Czerny \&
Grindlay 1986). Just below this luminosity level, its spectrum is
dominated by a much harder power-law component and the blackbody (BB) component
becomes almost undetectable above 2-3 keV (Czerny, Czerny \& Grindlay
1987; Harmon \etal\ 1996).  At even lower $L{x}$ ($\sim$10$^{33}$
ergs/s), its spectrum again is dominated by a BB component (Verbunt
\etal\ 1994).

In this paper, we report a clear spectral state transition in Aql X-1
detected with RXTE in February and March 1997 during an outburst decay.
We propose that the soft to hard state transition in Aql X-1 is due to
the centrifugal barrier, also commonly known as the `propeller'
effect.

\section{Observations and Results}

We followed the decay phase of the outburst of Aql X-1 in February and
March 1997 with RXTE (Bradt, Rothschild \& Swank 1993), as shown in
figure 1.  The hardness ratio (between the PCA count rates in 10-30 and
5-10 keV bands) remained remarkably stable during the first 17 days of
observations until near the end of the outburst decay between MJD 50510
and 50512 (March 5, 1997), when the hardness ratio increased suddenly.

We performed a spectral analysis for the last several days of
observations, as listed in table 1. Since the spectral model consisting
of a BB component plus a power-law component has been applied
successfully to the spectra of Aql X-1 (Czerny, Czerny \& Grindlay
1986) and other similar NS SXTs, for example, 4U1608-52 (Mitsuda
\etal\ 1984, 1989), we also adopted this spectral model to the RXTE/PCA
data. An iron line feature is also included.

It can be seen clearly that after around MJD 50510  (March 3, 1997)
(figure 1), the BB component luminosity within the total luminosity
decreased significantly and the power-law became much harder in the
meantime. The spectral state transition luminosity is between
(4-12)$\times$10$^{36}$, assuming a distance of 2.5 kpc. 
\section{A Unified Model for Soft X-ray Transients}

\subsection{Soft State or Regular Accretion State}

The radius of
the NS magnetosphere ($R_{m}$) varies slowly with the mass accretion rate
($\dot{M}$) onto the neutron star (Lamb, Pethick \& Pines 1973; Cui 1997): 
 
\begin{equation}
R_{m} = 10^{7}L_{x,36}^{-2/7}M_{1.4}^{1/7}B_{9}^{-4/7}R_{6}^{10/7} {\rm cm}
\end{equation}
where $L_{x,36}$ is the bolometric X-ray luminosity in 10$^{36}$ ergs/s,
$B_{9}$ is the NS magnetic field strength in
units of 10$^{9}$ G, 
$M_{1.4}$ is the NS mass in units of 1.4 \msol, and $R_{6}$ is its
radius in units of 10 km. 

When $\dot{M}$ is so high that $R_{m}$ is smaller than the last stable orbit
($3R_{S}$=$\frac{6GM}{c^{2}}$, where $M$ is the NS mass) of the NS, the
inner disk radius ($R_{in}$) is $3R_{S}$ (the recently detected
kilohertz QPOs imply that the NSs are indeed within their last stable
orbit; see for example Kaaret, Ford \& Chen 1997, Miller, Lamb \&
Psaltis 1996 and Zhang, Strohmayer \& Swank 1997).  In this state both
the inner accretion disk and the NS surface produce BB emissions with a
characteristic temperature of $\sim$ 1 keV (see, for example, Mitsuda
\etal\ 1989). A hard X-ray power-law component may be produced via
Comptonization of the BB emission photons by the bulk motion of the
relativistic inflow from the last stable orbit towards the NS surface
(Hanawa 1991, Kluzniak \& Wilson 1991, Walter 1992). Therefore its
continuum spectrum is made of a soft BB component plus a power-law.

The magnetosphere will expand when $\dot{M}$ decreases. When
$R_{m}>3R_{S}$, then $R_{in}=R_{m}$.  When $R_{m}>3R_{S}$ {\it and}
$R_{m}$ is smaller than the corotation radius ($R_{C}$), (almost) all
the material transferred from the companion falls onto the NS surface.
The overall spectral shape should remain basically unchanged. This
should correspond to the constant hardness ratio during the decay phase
of Aql X-1 outburst right before the sudden spectral transition.

In several similar NS systems, kilo-hertz OPOs have been detected previously. Such QPOs
can be explained reasonably well within the frame work of some kind 
of beat frequency model, in which the higher and lower frequency QPOs are the Keplerian
frequency of the inner region of the disk and its beat frequency against
the neutron star spin frequency, respectively (see van der Klis 1997 for a review).
Assuming that the higher frequency QPOs are
produced as a result of the interaction between the magnetosphere and
the inner accretion disk boundary, we would then expect that when the
magnetosphere is inside the peak emission radius of the inner accretion
disk region, no kilo-hertz QPOs are detectable. The peak radius is
roughly 1.4 times the radius of the last (marginal) stable orbit of the
NS (Zhang, Cui \& Chen 1997). Therefore $B$ can also be estimated from
the luminosity for the first appearance of its kilo-hertz QPOs, i.e,
the highest QPO frequency,  of the source during a luminosity decay,
using the relation $R_{m}=1.4\times 3R_{S}$. In a companion paper
(Zhang \etal\ 1997), the luminosity of Aql X-1 is reported to be
$\sim$2$\times$10$^{36}$ ergs/s when kilo-hertz QPOs were detected.
Thus the inferred strength of the NS magnetosphere is
$\sim$0.7$\times$10$^{8}$ G.

When $R_{C}>R_{m}>1.4\times 3R_{S}$, the QPO frequency and the X-ray
luminosity ($L{x}$) should be correlated positively. The exact
correlation depends upon the structure of the accretion disk just
outside the magnetosphere. When the vertical structure of the inner
accretion disk region changes, the relationship between the QPOs and
$\dot{M}$ will also change accordingly. Such evidence has been reported
by some of us for the same Aql X-1 outburst decay (Zhang \etal\ 1997).

\subsection{Hard State or Propeller State}

When $R_{m}>3R_{S}$ {\it and} $R_{m}>R_{C}$, the accreted material can
no longer overcome the centrifugal barrier, thus the mass accretion
onto the  surface will be reduced substantially. This is the well known
`propeller' effect (Illarionov \& Sunyaev 1975) Stella, White and Rosner (1986)
pointed out that `propeller' effects might be responsible for the transient
nature of many pulsars.  Such evidence has been reported in two X-ray 
pulsars GX1+4 and GROJ1744-28
(Cui 1997), when their pulsed emission disappeared suddenly during
their decay phases. The inferred values of $B$ are $>$10$^{13}$ G and
$\sim$2$\times$10$^{11}$ G for GX1+4 and GROJ1744-28, respectively.

When the propeller effects take place in a weakly magnetized
($B<10^{9}$ G) NS X-ray binary (the X-ray flux is usually not pulsed),
the flux of the BB component from the NS surface should decrease
significantly. In the meantime the BB component in the X-ray range
from the inner accretion disk edge should also decrease significantly,
since the inner disk region is disrupted by the NS magnetosphere.
Therefore the overall BB flux should decrease significantly.

The power-law component should become much harder in the meantime. One
reason is that the cooling effect of the BB photons from the NS surface
and the inner disk BB emission should become much weaker, since thermal
comptonization of these low energy photons by hot electrons has been
regarded as the primary mechanism for the hard X-ray power-law
production (see for example a recent review by Tavani and Barret (1997)
and references therein). The other is that the plasma just outside the
magnetosphere will become very hot in the propeller regime (Wang \&
Robertson 1985), thus will make the power-law component even harder.
As a consequence, the BB component will decrease suddenly and the
power-law will become much harder.

Additional hard X-ray emission may also be produced by
the relativistic outflow as a result of the propeller effects.  For a
1.8 or 3.6 ms NS, the Keplerian velocity at the corotation radius is
0.29$c$ and 0.23$c$, respectively. Thus the outflow is certainly
relativistic. Inverse Compton scattering of low energy photons by the
bulk motion of the outflow may also be responsible for the power-law
hard X-ray production, in a similar way as the suggested hard X-ray
production by the bulk motion of the accreted material {\it towards}
the NS (Hanawa 1991, Kluzniak \& Wilson 1991, Walter 1992).

We therefore identify the sharp spectral transition between MJD 50510
and 50512 is due to the expected propeller effects. The critical
luminosity corresponding to the appearance of the expected propeller
effect is (Lamb, Pethick \& Pines 1973; Cui 1997):  \
\begin{equation}
L_{x,36} \approx
2.34B_{9}^{2}P_{-2}^{-7/3}M_{1.4}^{-2/3}R_{6}^{5}
\end{equation}
where  $P_{-2}$ is the spin period in units of 10 ms. 
Assuming the NS mass of 1.4 \msol~  and a
neutron star radius of 10$^{6}$ cm, the inferred value of $B$ is thus
($1.5-2.5)\times 10^{8}$ G  or $(0.7-1.1)\times 10^{8}$ G for
$L_{x}=(4-12) \times 10^{35}$ ergs/s, corresponding to $P$=3.6 ms or
1.8 ms, respectively. Considering the consistency with the value of
$B$ estimated from the first appearance of QPOs, {\it the most likely
value of $B$ is $\sim$10$^{8}$ G and $P$=1.8 ms, i.e., the observed 549
Hz oscillation frequency is the NS spin frequency rather than one of
its harmonic frequencies}. We note that the uncertainty in estimating
$R_{m}$ (Wang 1996) does not affect the above consistency comparison, since
we applied the same equation from Lamb, Pethick \& Pines (1973) in both cases, 
so that
the uncertainty is cancelled out. Similar spectral state transitions were also
observed in Cen X-4 at a luminosity level of $\sim$8$\times$10$^{35}$
ergs/s (Bouchacourt \etal\ 1984). Then $B$$\sim$2$\times$10$^{9}$ G, if
$P$$\sim$33 ms (Mitsuda \etal\ 1996). In figure 2 we plot the
relationship between $B$ and $L_{x}$ for different values of $P$. The
locations of two X-ray pulsars (GX1+4 and GROJ1744-28, from Cui 1997)
and two soft X-ray transients (Aql X-1 and Cen X-4) are also marked.

The exponential decay constant ($e$-folding time) of the outburst is
$\sim$20 days. The amount of  decay expected in two days would have
caused a luminosity decrease by $\sim$10\%. Therefore the observed
sudden BB luminosity decrease implies that $\ga$90\% of the material is
`propelled' out and only $\la$10\% of the material reached the NS
surface. We note here that whether one uses the total flux or just
the BB flux to estimate the change in $\dot{M}$ after the spectral transition
does not qualitatively change this conclusion.

It is also interesting to check the consistency of our above
interpretation with the model of Wang and Robertson (1985). Using
$B$=10$^{8}$ G and $\dot{M}$=10$^{-12}$ \msol/year (corresponding to
10$^{36}$ ergs/s for a 20\% rest mass conversion efficiency) and taking
all three factors used in their equations (30), (33) and (35) to be
unity, we obtain the equilibrium  period of the NS to be $\sim$2 ms,
the mean plasma temperature just outside the boundary layer of the
magnetosphere to be $\sim$2$\times$10$^{9}$ K ($kT\sim$170 keV), and
the maximum energy of the non-thermal particles produced within the
neutral sheets located at the boundaries of the vortex structures to be
$\sim$4$\times$10$^{15}$ eV. It is interesting to note that for a
similar spectrum in a similar system 4U1608-52, the inferred $kT$ from
an observed broken power-law is $\sim$65 keV (Zhang \etal\ 1996). The
possibility of producing high energy particles implies that SXTs in the
propeller phase may be gamma-ray emitters.

In this state we also expect that the kilo-Hertz QPOs will cease to
exist or become much weaker. This could be an additional indicator of
the on-set of the propeller phase. In fact for Aql X-1, high frequency
QPOs became undetectable just before the observed spectral state
transition, or the propeller effects take place.

\subsection{Quiescent or Advection Dominated States}

When $\dot{M}$ decreases further, the disk itself becomes advection
dominated and its entire optically thick inner region is truncated
(Narayan 1996; Hameury \etal\ 1997).  The radius at which the rotation
velocity of the NS magnetosphere reaches the speed of light is
$\sim$10$^{7-8}$ cm for millisecond spinning NSs, still very close to
the NS surface. Thus the inner boundary of an advection dominated disk
($>$10$^{9-10}$ cm) is very far away from the NS magnetosphere. In this
case, the mass transfer from the outer accretion disk onto the NS
should be nearly spherical.  Therefore some infalling matter can escape
the centrifugal barrier and falls onto the polar regions of the NS to
produce the observed weak BB luminosity.  The inferred small radii of
the BB emission regions in the quiescent states of several SXTs
(Verbunt \etal\ 1994; Asai \etal\ 1996) also agree with this picture.
In some cases, X-ray pulsations are expected, if the NS spin and the
magnetic field axis are mis-aligned and that X-ray emission regions
cover only part of the NS surface. Indeed weak X-ray pulsations were
reported in a similar system
 Cen X-4 during its quiescent state (Mitsuda \etal\ 1996). 
 It is also interesting to note
 that Cen X-4 is so far the only NS SXT possibly observed with X-ray
 pulsations in the quiescent state, and yet its inferred value of $B$
$\sim$2$\times$10$^{9}$ G may be near the upper end of the expected
value for type-I X-ray bursters (see for example, Joss 1978; Taam \&
Picklum 1978; Joss \& Li 1980).  A hard power-law component may also
be produced, again
due to the relativistic outflows.  In this case the seed photons
available for inverse Compton scattering are much less populous than
when the system just enters the propeller state, since the advection
dominated outer disk is cold and does not produce much radiation.

In our model the propeller and quiescent state properties of SXTs are
different from previous models of Stella \etal\ 1994 and Verbunt
\etal\ 1994. Verbunt \etal\ (1994) argued that a detection of any 
significant X-ray flux
(more than the level possible from the low mass companion star) in a
X-ray binary system implies that the system is {\it not} in its
propeller regime. In the model of Stella \etal\ (1994), accretion onto 
the neutron star surface is also not possible in the propeller regime, but UV or soft X-ray 
emission is possible at the magnetospheric radius. In our model, 
X-ray emission and the observed
spectral shape in the propeller or an even lower accretion rate state
-- the quiescent state are very natural. We emphasize that the major difference
between our model and the previous ones is that {\it in the quiescent state
of our model,
the disk is truncated, due to the advection dominance, very far away from the 
light cylinder
of the neutron star magnetosphere}. Thus near-spherical accretion onto the neutron star
poles are possible. We predict that {\it some SXTs in quiescence should be millisecond
X-ray pulsars}.

As this paper is ready for submission, we learnt that another paper by
Campana \etal\ (1997) is near its completion for submission. In that
independent paper, Campana \etal\ report a rapid luminosity decrease of
Aql X-1 around and after the spectral state transition reported in this
paper, from the SAX/NFIs observations during the same outburst decay.
They arrived at almost identical conclusions concerning the transition
to ours in this paper.

{\it Acknowledgement: }
SNZ appreciates deeply the many stimulating discussions with Wei Cui,
Wan Chen and Marco Tavani, which initially brought his attention to
this topic. The anonymous referee is highly appraised for his/her
prompt response and providing many valuable suggestions and
constructive criticism, which certainly improved the overall quality of
this paper. We also thank Didier Barret, Luigi Stella and Marco Tavani for
making useful comments on the manuscript. 
SNZ also acknowledges partial financial support by NASA
through contracts NAG5-3681, 4411 and 4423.

\small
\begin{table}
\begin{tabular}{cccccccccc}
\hline
Obs. & \multicolumn{2}{c}{Power Law}& \multicolumn{2}{c}{Blackbody} & 
\multicolumn{2}{c}{6.4 keV Iron Line } & Total & $\frac{\rm BB}{\rm Total}$ & 
$\frac{\chi^{2}}{39}$ \\
& Index & Norm & kT$_{bb}$ & L$_{bb}$ &  1-$\sigma$ Width & Flux & L$_{x}$  &&\\
(1) & (2) & (3) & (4) & (5)& (6) & (7)& (8) & (9)& (10)\\
\hline

09 &	
3.26$\pm$.01& 
1.96$\pm$.04& 
1.51$\pm$.01 &  
5.37$\pm$.04 &
.59$\pm$.05 &
11.$\pm$1.&
1.47 & .37& 
2.49 \\

10 &
2.29$\pm$.02 &
.20$\pm$.01 &
1.17$\pm$.04 &
.43$\pm$.02 &
.87$\pm$.05 &
10.$\pm$1.&
0.45 & .10 &
1.37 \\

11 & 
1.83$\pm$.03&
.028$\pm$.002&
.50$\pm$.03&
.32$\pm$.04&
.33$\pm$.20&
.80$\pm$.30 &
0.15 & .21 &
1.10 \\
\hline
\end{tabular}
\caption{Model spectral fitting results for the last three observations shown in figure 1.
PCA data between 3-20 keV are used ($N_{H}=3.3E21/cm^{2}$).
{\it Notes:} (1) No. 09: MJD 50509.906--50510.025, No. 10: 50511.951--50512.062, No. 11: 
50514.892--50515.055; (2) Photon spectral index; (3) photon/keV/cm$^{2}$/s at 1 keV;
(4) keV; (5) 10$^{35}$ ergs/s at 2.5 kpc; (6) keV; (7) 10$^{-4}$ photon/cm$^{2}$/s in the line;
(8) 1.5-30 keV, 10$^{36}$ ergs/s at 2.5 kpc.}
\end{table}
\normalsize

\begin{figure}
\centering{
\epsfig{figure=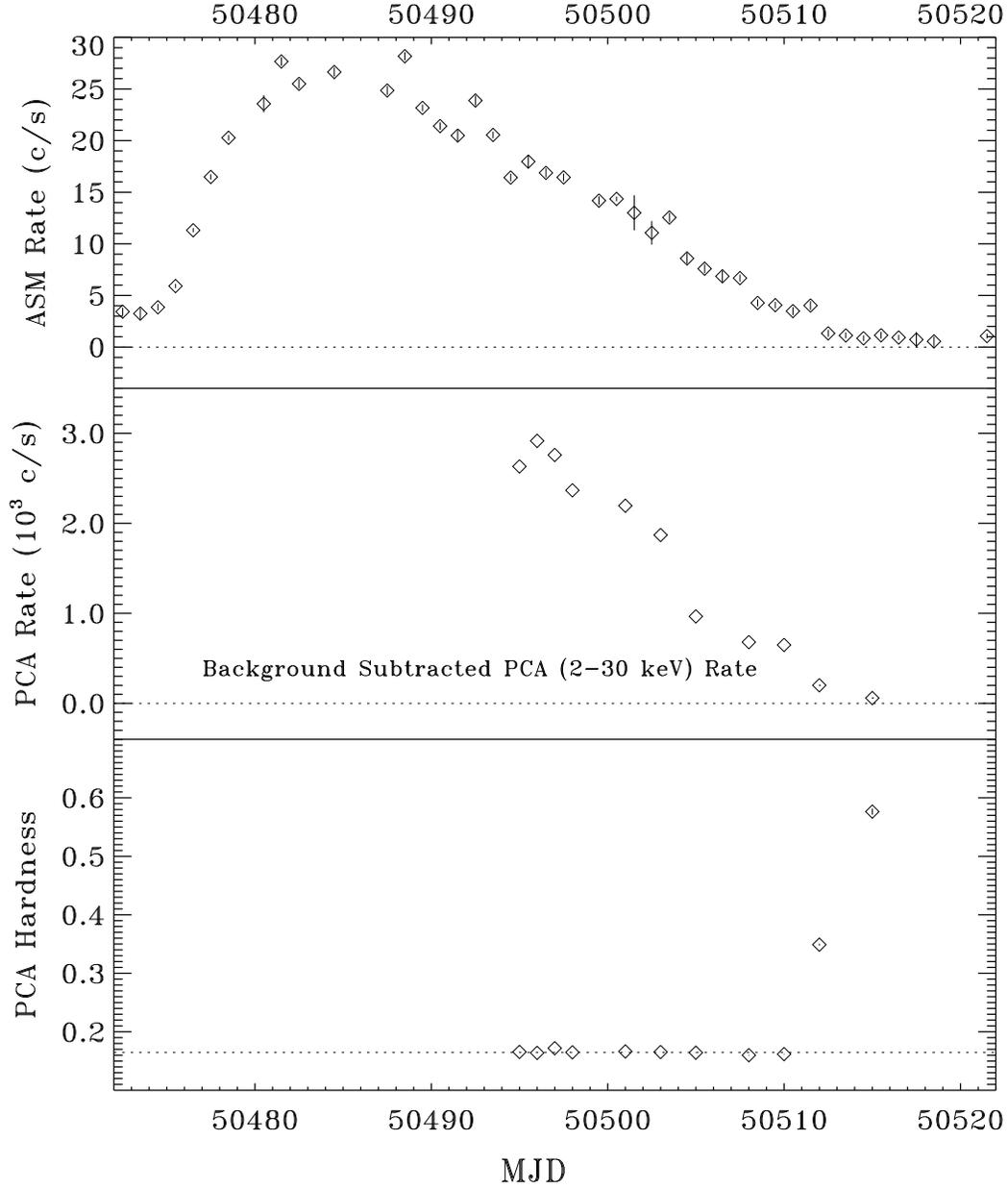,width=6.0in}
}
\caption{Spectral evolution of Aql X-1 during its outburst decay phase. The hardness ratio
(10-30/5-10 keV) remained as almost a constant when the PCA count rate decreased from
about 3000 c/s to about 700 c/s and then suddenly increased significantly when the
count rate was around and below 200 c/s.}
\end{figure}
\begin{figure}
\centering{
\epsfig{figure=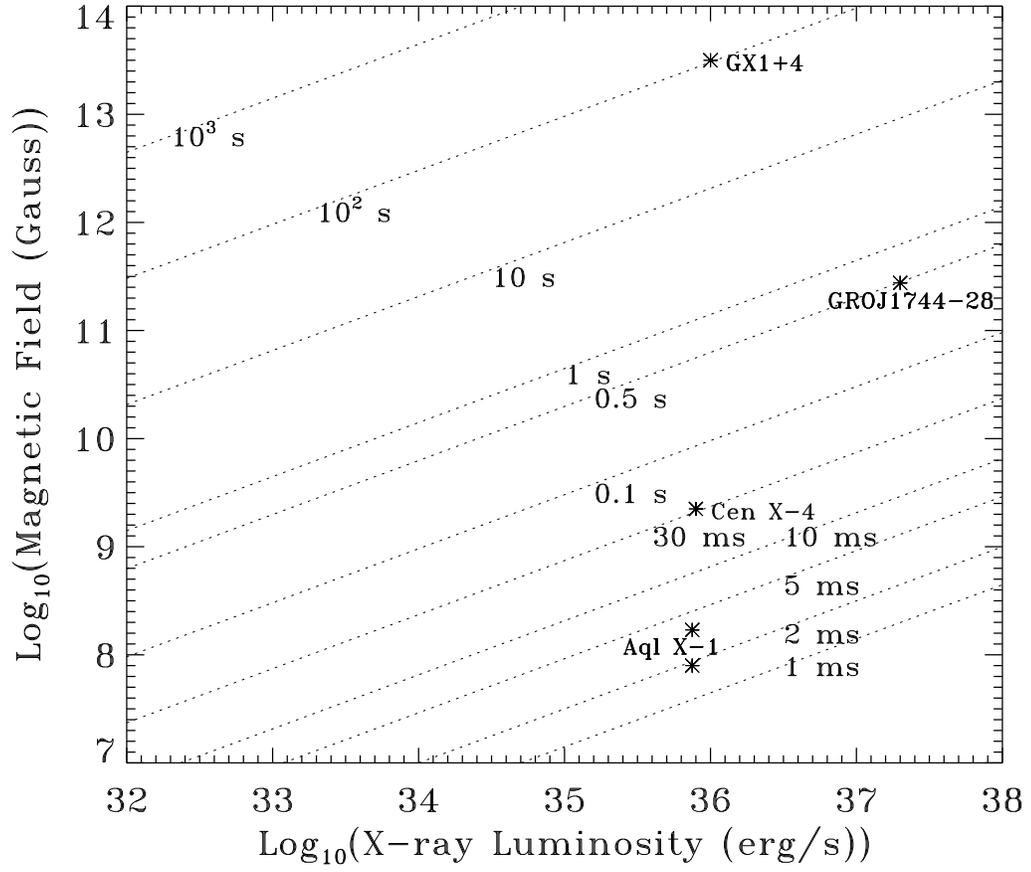,width=6.0in}
}
\caption{Relationship between the neutron star magnetic field strength and the critical
luminosity level at which the propeller effect takes place. The locations of GX1+4,
GROJ1744-28, Aql X-1 and Cen X-4 are also marked.}
\end{figure}
\end{document}